\documentclass{svjour2}   
\smartqed 
\usepackage{graphicx}

\begin{document}

 \title{Energy fluctuations shape free energy of nonspecific biomolecular interactions} 
\author {Michael Elkin \and Ingemar Andre \and David B. Lukatsky}
\institute{D. B. Lukatsky \at 
Department of Chemistry,  Ben-Gurion University of the Negev, 84105 Beer-Sheva, Israel \\
Tel.: +972-8-642-8370\\
Fax: +972-8-647-2943\\
\email{lukatsky@bgu.ac.il}
\and 
M. Elkin \at 
Department of Computer Science, Ben-Gurion University of the Negev, 84105 Beer-Sheva, Israel 
\and
I. Andre \at 
Department of Biochemistry and Structural Biology, Lund University, 221 00 Lund, Sweden}

\date{Received: date / Accepted: date}

\maketitle

\begin{abstract}
Understanding design principles of biomolecular recognition is a key question of molecular biology. 
Yet the enormous complexity and diversity of biological molecules hamper the efforts to gain a predictive ability
for the free energy of protein-protein, protein-DNA, and protein-RNA binding. 
Here, using a variant of the Derrida model, we predict that for a large class of biomolecular interactions, it is possible to accurately estimate the relative free energy 
of binding based on the fluctuation properties of their energy spectra, even if a finite number
of the energy levels is known. 
We show that the free energy of the system possessing a wider binding energy spectrum is almost surely lower compared
with the system possessing a narrower energy spectrum.  
Our predictions imply
that low-affinity binding scores, usually wasted in protein-protein and protein-DNA docking algorithms, can be efficiently utilized to compute the free energy.
Using the results of Rosetta docking simulations of protein-protein interactions from Andre et al., {\it{Proc. Natl. Acad. Sci. U.S.A.}} {\bf{105}}, 16148 (2008), we demonstrate the power of our predictions.
\end{abstract}

\def\Var{\mathit{VAR}}
\def\hY{\hat{Y}}
\def\hsigma{\hat{\sigma}}

\section{Introduction}

Recent high-throughput experiments demonstrate a high level of multi-specific and non-specific binding in protein-protein~\cite{ppi}, protein-DNA \cite{polly}, and protein-RNA~\cite{rna} interactions in a living cell.
These observations challenge a conventional approach of molecular biology usually focusing on just a single pathway or function, or a single binding partner for a protein.
This suggests that in order to predict correctly the properties of molecular interaction networks, one needs to take 
into account the effect of multiple binding, essentially computing the free energy of the system rather than the energy of individual states. 
The latter statement is quite intuitive as any protein in a cell interacts with thousands of proteins (or DNA binding cites), and even if one (or few) of its interaction 
partners have stronger binding affinity than others, still weaker interactions are not negligible and they may become even dominant.  
Yet the complexity of biological molecules and a lack of knowledge of accurate inter-molecular interaction potentials hamper
computational efforts to predict the free energies of protein-protein, protein-DNA, and protein-RNA binding. 
A key question is how to estimate the binding free energy based on the partial knowledge of the binding energy spectrum. Each energy in the binding energy spectrum
is defined here as the inter-molecular interaction energy of a particular bound state  of interacting molecules (e.g., particular binding configuration of protein-protein or protein-DNA complex). 

It was recently shown that global symmetry properties of proteins, both on structural and sequence levels, generically define the properties of their binding energy spectrum \cite{lukatsky_prl,lukatsky_jmb,baker_pnas,lukatsky_pre,lukatsky_ariel,lukatsky_jmb_2011}. 
In particular, it was shown analytically in~\cite{lukatsky_prl} that the probability distribution for the interaction energies of homodimers, $P(E)$, is 
always wider as compared to heterodimers, $\sigma_\mathrm{homo}/\sigma_\mathrm{hetero}=\sqrt{2}$, where $\sigma$ is the dispersion of $P(E)$. 
This statistical law was also confirmed computationally, using one of the 
most advanced methods for computing protein-protein interactions applied to a large dataset of protein complexes from the protein data bank (PDB) \cite{baker_pnas}.
It was also predicted that proteins possessing a higher level of structural correlations (clustering) of amino acids in their interfaces,
demonstrate a wider binding energy spectrum, as well~\cite{lukatsky_pre}. It was shown recently that protein sequences with enhanced  
strength of diagonal correlations of amino acid positions demonstrate a similar property \cite{lukatsky_ariel,lukatsky_jmb_2011}. Intuitively it means that the clustering of amino acids of the same type
statistically enhances the dispersion of the binding energy spectrum~\cite{lukatsky_ariel,lukatsky_jmb_2011}. Sequences with a higher 
level of such clustering will possess a larger dispersion than sequences with a lower level of clustering ~\cite{lukatsky_ariel,lukatsky_jmb_2011}. 
We have recently analyzed the properties of the energy spectrum of nonspecific protein-DNA binding~\cite{sela_2011}. Similar to the case of protein-protein interactions, we also observed that the width of the protein-DNA binding energy spectrum depends on the correlation properties of DNA, such as
the symmetry and the length-scale of DNA sequence correlations~\cite{sela_2011}.

\begin{figure}
\begin{center} 
\includegraphics[width=11cm]{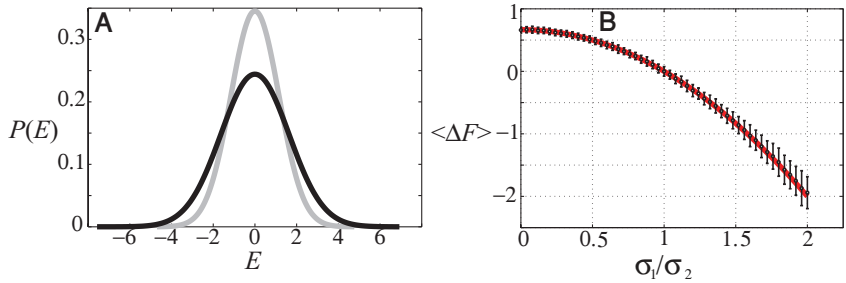}
\end{center} 
\caption{Calculation of the free energy and fluctuations of the free energy from the energy spectrum. {\bf{A.}} Example: Gaussian probability distributions for the interaction energy, $E$, characterized by the identical average energies, $\left<E_1\right>=\left<E_2\right>$, and different
dispersions, $\sigma_1/\sigma_2=\sqrt{2}$. $E$ is represented in the units of $k_BT$.  
{\bf{B.}} Computed average free energy differences, $\left<\Delta F\right>=\left<F(\sigma_1)-F(\sigma_2)\right>$, as a function of the ratio of dispersions. Circles with error bars represent the simulation results, where the quenched averaging is performed (see the text). We used $M=1000$ 
for each computation of the partition function, and the averaging is performed with respect to $200$ realizations. $\left<\Delta F\right>$ is represented in the units of $k_BT$. Error bars represent free energy fluctuations, and show two standard deviations. Solid curve represents the analytical result, Eq. (\ref{eq:free_en}). \label{fig_gaussian}}
\end{figure}

We emphasize that in all of those examples 
the average interaction energies of the compared spectra are always identical, and only the dispersions of the energy spectra are different, Fig. \ref{fig_gaussian}. 
The predicted effects are thus essentially governed by the fluctuations of energy, 
and go beyond the mean field. The case where the average energies are not equal is also discussed below. 
We assume here that the probability distribution, $P(E)$, is Gaussian. This is an accurate assertion since, practically, the binding energy, $E$, is a sum
of thousands of binary inter-atomic interactions, and this sum is normally distributed according the the central limit theorem~\cite{lukatsky_prl}. 

Here, we estimate the relative free energy of two interacting systems characterized by the same average binding energies, $\left<E_1\right>=\left<E_2\right>$, but different dispersions, $\sigma_1>\sigma_2$, Fig. \ref{fig_gaussian}A. We show that the free energy, $F$, of the system possessing a wider binding energy spectrum, is always shifted towards lower free energies compared to the system possessing a narrower $P(E)$, even if a finite number of energy levels is known, Fig. \ref{fig_gaussian}A. 
In particular, we show that the partition function, $Z_1$, is almost surely larger than $Z_2$,  with the probability, $P(Z_1>Z_2)\geq 1-C/M$, where $C(\sigma_1, \sigma_2)$ is a finite constant, and $M$ is the number of the energy levels used to compute the partition function. 

We note that in his seminal work \cite{derrida}, Derrida has established that in the random energy model, where the energy spectrum of the system, $P(E)$, is Gaussian, and the partition function, $Z=\sum_{i=1}^{M} \exp(-E^{(i)}/k_BT)$, for each realization of $P(E)$ with $M$ energy states, $E^{(i)}$ , the  quenched average of the free energy, $\left<F\right>_q=-k_BT\left<\ln Z\right>$, is equal to the annealed average,   $\left<F\right>=-k_BT\ln\left< Z\right>$, in the thermodynamic limit of large $M$:
\begin{equation}
\label{eq:derrida}
\left<F\right>=-k_B T\ln M-\frac{\sigma^2}{2k_BT},
\end{equation}
if the temperature $T$ is above some critical temperature, $T>T_c$, where $k_BT_c\sim \sigma /\sqrt{\ln M}$ \cite{derrida}, and $\sigma$ is the standard deviation of $P(E)$. Using an example from the Rosetta docking simulations of protein-protein interactions, we show below that Eq. (\ref{eq:derrida}) provides an accurate
estimate for the relative free energy, when $M$ reaches only few thousands.

We stress that our model is applicable to interacting systems without a pronounced low-energy (ground) state in their energy spectra. The existence of such a ground state corresponds to a strong, specific binding. On the contrary, a large class of weakly interacting biomolecules in a living cell, such as nonspecific protein-protein, protein-DNA, or protein-RNA binding, represents the systems where our model is operational. Such relatively weak, nonspecific interactions, often called ``promiscuous interactions'', have been
shown to play an important role in different cellular processes, and in many cases, the effect of such weak interactions becomes the dominant factor in a living cell \cite{tawfik}.

\section{Results}

We consider the ensemble of the interaction energies, $\{E^{(i)}\}$, of two interacting biomolecules, where each energy, $E^{(i)}$, corresponds to
a given conformation (i.e., a given interaction state), $i$, of these molecules. 

We begin with the definition of the free energy of the system, $F=-\ln Z$, where the partition function, $Z=\sum_{i=1}^M e^{-E^{(i)}}$, 
and we assume for simplicity that $k_BT=1$, and the energy, $E$, is represented in the units of $k_BT$; here $k_B$ is the Boltzmann constant, and $T$
is the absolute temperature. 
We also assume that the energy, $E$, obeys the Gaussian distribution, $P(E)$, with zero mean, $\left<E\right>=0$, and standard deviation, $\sigma$. The set of $M$ energy values, 
$E^{(i)}$, is obtained as a statistical realization of $P(E)$. 
In what follows we compare the statistical properties of $Z$ computed based on the realizations drawn from two distributions with different values of standard deviation, $\sigma_1>\sigma_2$, Fig. \ref{fig_gaussian}A.

Since $Y = e^{-E}$ is a lognormal random variable,
it is well-known~\cite{AB57} and can be readily verified that the
expectation, $\left<Y\right>$, and the variance, $\Var(Y)$, of $Y$ are given by
$\left< Y \right> = e^{\sigma^2/2}$ and $\Var(Y) = e^{2\sigma^2} - e^{\sigma^2}$. We note that Eq. (\ref{eq:derrida}) simply follows from
$\left<F\right>=-\ln\left( M \left< Y \right>\right)$. 

For a large number $M$, let $Y^{(1)},Y^{(2)},\ldots,Y^{(M)}$ be
independent random variables, so that each of them is distributed
identically to $Y$. Since $Z = \sum_{i=1}^M Y^{(i)}$,
by linearity of the expectation, $\left< Z \right> = M \cdot e^{\sigma^2/2}$.
Also, since $Y^{(1)},Y^{(2)},\ldots,Y^{(M)}$ are independent, it follows
that
\begin{equation}
\label{eq:variance}
\Var(Z) = \Var(\sum_{i=1}^M Y^{(i)}) = \sum_{i=1}^M \Var(Y^{(i)}) = 
M \cdot (e^{2\sigma^2} - e^{\sigma^2})~,
\end{equation}
and the standard deviation of $Z$, $\sigma(Z) = \sqrt{\Var(Z)}$.

Consider now two normal independent random variables $E_1$ and $E_2$,
both having zero mean.
We assume further that the standard deviation $\sigma_1$ of $E_1$ is greater
than
the standard deviation $\sigma_2$ of $E_2$, i.e., $\sigma_1 > \sigma_2 > 0$.
Let $Y_1 = e^{-E_1}$ and $Y_2 = e^{-E_2}$ be the corresponding lognormal random
variables. 
Next we show that asymptotically almost surely it holds that $Z_1 >Z_2$, where $Z_1= \sum_{i=1}^M Y_1^{(i)}$
and  $Z_2= \sum_{i=1}^M Y_2^{(i)}$.

\begin{lemma}
Let $\sigma_1,\sigma_2$, $\sigma_1 > \sigma_2 >0$, be two positive
constants. Then
\begin{equation}
\label{eq:prob}
P(Z_1 > Z_2) ~\ge~ 1 - {1 \over M} \cdot f(\sigma_1,\sigma_2)~,
\end{equation}
where $$f(\sigma_1,\sigma_2) ~=~ 4\cdot {{(e^{2\sigma_1^2} - e^{\sigma_1^2})
+
(e^{2\sigma_2^2} - e^{\sigma_2^2})} \over
{(e^{\sigma_1^2/2}-e^{\sigma_2^2/2})^2}}$$ 
is a positive constant that depends only on $\sigma_1$ and $\sigma_2$.
\end{lemma}

\begin{proof}
Chebyshev's inequality (see, e.g.,~\cite{AS08}, p.43) states that
for any random variable $X$ with expectation $\left<X\right>$ and standard
deviation $\sigma_X$, and any $b > 0$,
\begin{equation}
\label{eq:Cheb}
P(|X - \left<X\right>| \ge b \cdot \sigma_X) ~\le~ {1 \over {b^2}}~.
\end{equation}
By the preceding argument (see Eq. (\ref{eq:variance})), for
$i=1,2$,
\begin{eqnarray*}
&&\left< Z_i\right> ~=~ M \cdot e^{\sigma_i^2/2},~
\sigma (Z_i) ~=~ \sqrt{M} \cdot \sqrt{e^{2 \sigma_i^2} -
e^{\sigma_i^2}}.
\end{eqnarray*} 
Let $$A ~=~ {{\left< Z_1 \right> + \left< Z_2 \right>} \over 2} ~=~ 
M \cdot {{e^{\sigma_1^2/2} + e^{\sigma_2^2/2}} \over 2}~.$$
Hence 
\begin{equation}
\label{eq:diff}
\left< Z_1 \right> - A ~=~ A - \left< Z_2 \right> ~=~  {{\left< Z_1 \right> -
\left< Z_2 \right>} \over 2} ~=~ M \cdot
 {{e^{\sigma_1^2/2} - e^{\sigma_2^2/2}} \over 2}~.
\end{equation}
Denote $Q = {{e^{\sigma_1^2/2} - e^{\sigma_2^2/2}} \over 2}$ and $D = M
\cdot Q$. 
Observe that since $\sigma_1 > \sigma_2$, both $D$ and $Q$ are positive.
Consequently, $\left< Z_1 \right> - D =  \left< Z_2 \right> + D = A$.
Also, by Chebyshev's inequality (see Eq. (\ref{eq:Cheb})),
\begin{eqnarray*}
P(|Z_1 - \left< Z_1 \right>| \ge D) = P(|Z_1 - \left< Z_1 \right>| \ge
\sqrt{M} \cdot {Q \over {\sqrt{e^{2\sigma_1^2} - e^{\sigma_1^2}}}} \cdot 
\sigma(Z_1) ) \\ 
~\le~ {1 \over M} \cdot {{e^{2\sigma_1^2} - e^{\sigma_1^2}} \over
{Q^2}}~.
\end{eqnarray*}

It follows that 
\begin{eqnarray*}
P(Z_1 \le A) &=& P(Z_1 \le \left< Z_1 \right> - D) = 
P((\left< Z_1 \right> - Z_1) \ge D) \\
& \le& 
P(|\left< Z_1 \right> - Z_1| \ge D) ~\le~  {1 \over M} \cdot
 {{e^{2\sigma_1^2} - e^{\sigma_1^2}} \over
{Q^2}}~.
\end{eqnarray*}
Analogously, 
$$P(Z_2 \ge A)  = 
P((Z_2 -\left< Z_2 \right>) \ge D) 
~ \le~
P(|Z_2 -\left< Z_2 \right>| \ge D) ~\le~  {1 \over M} \cdot
 {{e^{2\sigma_2^2} - e^{\sigma_2^2}} \over
{Q^2}}~.
$$
Hence, by union bound,
$$P((Z_1 > A) \mbox{~and~}
  (Z_2 < A)) ~\ge~ 1 - {1 \over M} \cdot {1 \over
{Q^2}} \cdot ((e^{2 \sigma_1^2} - e^{\sigma_1^2}) + 
(e^{2 \sigma_2^2} - e^{\sigma_2^2}))~.$$
Finally,
\begin{eqnarray}
P(Z_1 > Z_2) &\ge& P((Z_1 > A) \mbox{~and~} (Z_2 < A)) \nonumber \\ 
&\ge& 1 - {1 \over M} \cdot {1 \over
{Q^2}} \cdot ((e^{2 \sigma_1^2} - e^{\sigma_1^2}) + 
(e^{2 \sigma_2^2} - e^{\sigma_2^2}))~.
\label{eq:misha}
\end{eqnarray}
\end{proof}

Since the right-hand side of the inequality
(Eq. (\ref{eq:prob})) tends to 1 as $M$ grows, it follows that the event
$Z_1 > Z_2$ occurs asymptotically almost surely.
This argument generalizes directly to the scenario when
$Z_1 = \sum_{i=1}^{M_1} Y_1^{(i)}$
and 
$Z_2 = \sum_{i=1}^{M_2} Y_2^{(i)}$, where $M_1$ and $M_2$ are (not
necessarily equal) large integers. The generalized inequality is
\begin{eqnarray*}
P(Z_1/M_1 > Z_2/M_2) ~&\ge&~ 1 - {{4} \over {(e^{\sigma_1^2/2} -
e^{\sigma_2^2/2})^2}} \times \\ 
&\times&\left( {1 \over {M_1}} \cdot (e^{2\sigma_1^2} - 
e^{\sigma_1^2}) +  {1 \over {M_2}} \cdot (e^{2\sigma_2^2} - 
e^{\sigma_2^2})\right)~.
\end{eqnarray*}
It is easy to understand the obtained results intuitively. In the calculation of the partition function,
$Z=\sum_{i=1}^M e^{-E^{(i)}}$, $M$ energies $E^{(i)}$ are drawn from the Gaussian distribution. 
However, only a subset of {\it{lowest}} energies provides the dominant contribution to $Z$. The
contribution from high energies is small, $e^{-|E|}\ll 1$. 
Since this dominant subset is localized in the low energy tail, the distribution with a larger standard deviation 
will obviously deliver the larger partition function. 

The major practical implication of our result is the ability to estimate 
the relative free energy of biomolecular interactions without 
performing the actual calculations of the free energy. 
We establish that a simple, direct relationship between the average free energy difference and 
the standard deviations of the energy spectra, Eq. (\ref{eq:derrida}),
\begin{equation}
\label{eq:free_en}
 \left<\Delta F\right>=\left<F(\sigma_1)-F(\sigma_2)\right> =-\frac{\sigma_1^2-\sigma_2^2}{2},
\end{equation}
is accurate for a system where only a finite number of energy levels is known. We note that our analytical definition of the average free energy relies on the 
{\it{annealed}} definition of the average, $\left<F\right>=-\ln \left< Z \right> $. In systems without frustration the latter 
definition is known to be in excellent agreement with the {\it{quenched}} averaging, $\left<F\right>_q=- \left< \ln Z \right>$, unlike the case of highly frustrated systems such as spin glasses \cite{derrida} or proteins below the glass transition temperature ~\cite{shakh}. Indeed, the quenched averaging performed numerically is in excellent agreement with the analytical result, Fig. \ref{fig_gaussian}B. The error bars in this plot represent the magnitude of the free energy fluctuations. Yet, our central result in this paper is stronger than the statement described by Eq. (\ref{eq:free_en}).
Here we predict for two systems, that even if a single calculation of the free energy is performed for each system,  using a single realization of the probability distributions, $P(E_1)$ and $P(E_2)$, and it is known that 
$\sigma_1>\sigma_2$, then we guarantee that $F_1<F_2$ with the probability approaching one, provided that the number of
measured energy levels, $M$, in each realization is sufficiently large. Finally we note that if one of the distributions, $P(E_1)$, 
is shifted from zero mean by the energy, $\left<E_1\right>=E_0$, the free energy, Eq. (\ref{eq:free_en}) gets trivially shifted exactly
by this magnitude, $\left<\Delta F\right>=E_0 -(\sigma_1^2-\sigma_2^2)/2$. This is because the fluctuation contribution to the free energy difference depends exclusively
on the widths of the corresponding energy spectra.

\begin{figure}
\begin{center} 
\includegraphics[width=11cm]{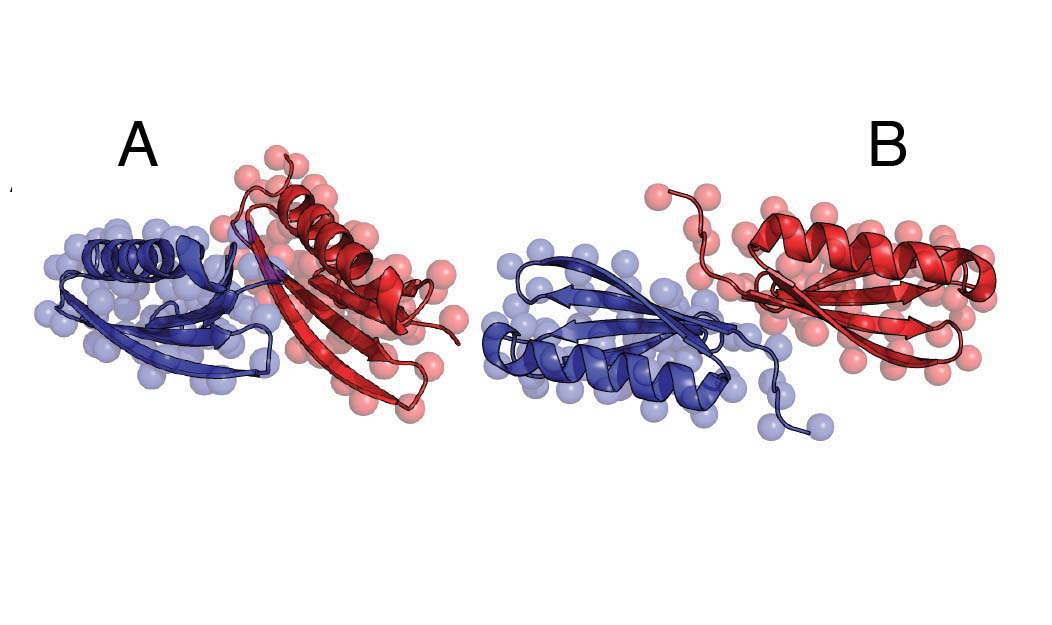}
\end{center} 
\caption{Snapshot of assymetric (we use the term ``heterodimeric'' to describe such symmetry) ({\bf{A}}), and symmetric (we use the term ``homodimeric'' to describe such symmetry) ({\bf{B}}) binding modes from Rosetta docking simulations of protein L (PDB code: 1hz6). The structures represent the lowest energy binding modes (using the same energy term as in ref. \cite{baker_pnas}, the interchain pair potential in the Rosetta low resolution docking energy function) from an assymetric or symmetric docking simulation of protein L dimers.  Position of centroid atoms, representing the sidechain atoms, is shown as spheres.   
\label{fig_rosetta}}
\end{figure}

\begin{figure}
\begin{center} 
\includegraphics[width=11cm]{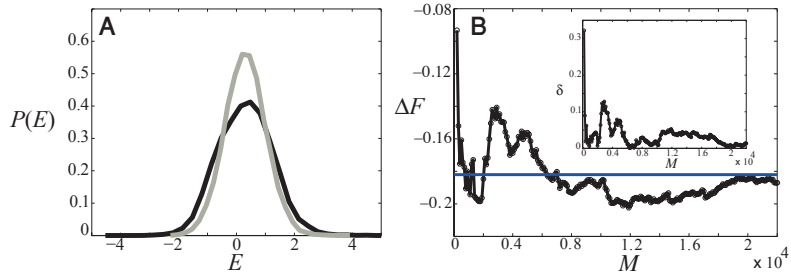}
\end{center} 
\caption{Example: Calculation of the free energy of nonspecific self-binding for protein L, using the Rosetta docking scores obtained by Andre et al. in ref. \cite{baker_pnas}.  {\bf{A.}} Computed probability distributions, $P(E)$, of the Rosetta docking energies for protein L in symmetric (i.e. homodimeric, shown in black) and nonsymmetric (i.e. heterodimeric, shown in gray) conformations \cite{baker_pnas}. Snapshots from these Rosetta docking simulations are shown in Fig. \ref{fig_rosetta}. The interaction energy, $E$, is in dimensionless Rosetta score units. There are overall 29,976 homodimeric,$\{E^{(i)}_{homo}\}$, and 22,038 heterodimeric, $\{E^{(i)}_{hetero}\}$, conformations sampled, respectively.  {\bf{B.}} Computed free energy difference between homodimers and heterodimers,  $\Delta F=F_{homo}-F_{hetero}$, as a function of the energy sample size, $M$, where, $F_{homo}=\sum_{i=1}^M \exp(-E^{(i)}_{homo})$, and analogously for $F_{hetero}$. The horizontal line represents the expectation value, $\left<\Delta F\right>=-(\sigma_{homo}^2-\sigma_{hetero}^2)/2$, Eq. (\ref{eq:free_en}), where $\sigma_{homo}$ and $\sigma_{hetero}$ are the standard  deviations of the corresponding energy spectra. Inset represents the relative deviation of $\Delta F$ from the expectation value, 
$\delta=|\Delta F-\left<\Delta F\right>|/|\Delta F+\left<\Delta F\right>|$, as a function of $M$.
\label{fig_ingemar}}
\end{figure}

\section{Example: Free energy of nonspecific protein-protein interaction}

We now apply our results to the calculation of the free energy of nonspecific protein-protein interactions. We use an example from Andre et al. \cite{baker_pnas},
where the Rosetta docking simulations of self-interacting protein L in homodimeric and heterodimeric conformations were performed \cite{baker_pnas} (see Fig. \ref{fig_rosetta}).
In particular, these simulations provide the interaction energies of $\sim30,000$ homodimeric and $\sim22,000$ heterodimeric conformations, respectively Fig. \ref{fig_ingemar}A. Each of these conformations is chosen randomly, without any energy optimization. Based on these energies, we computed the free energy difference $\Delta F=F_{homo}-F_{hetero}$, as a function of the energy sample size, $M$, where, $F_{homo}=\sum_{i=1}^M \exp(-E^{(i)}_{homo})$, and analogously for $F_{hetero}$ (see Fig. \ref{fig_ingemar}B). The key result here is that the free energy of homodimeric conformations is always lower then the corresponding free energy of heterodimeric conformations, Fig. \ref{fig_ingemar}B. After the sample size, $M$, reaches only few thousands conformations, the
free energy difference reaches its expectation value, $\left<\Delta F\right>$, with the accuracy reaching 90-95\% (see inset in Fig. \ref{fig_ingemar}B). The estimate
performed above, Eq. (\ref{eq:misha}), gives: $P(Z_{homo}>Z_{hetero}) \geq 0.95$, if $M=5000$, where $Z_{homo}$ and $Z_{hetero}$ are the corresponding partition functions,  each obtained based on $M$ energy values. We suggest therefore that our method should provide an efficient way to estimate
the free energies of nonspecific biomolecular binding.

\section{Conclusion}

The majority of macromolecular docking algorithms rejects the lower-affinity binding scores and retain only one or few  
lowest energy conformations. We suggested here a simple method based on Derrida-type random energy model \cite{derrida}, how those wasted scores can be used in order to estimate 
the free energy of binding. Our conclusions can be applicable to different biomolecular systems, such as protein-protein, protein-RNA, and protein-DNA
complexes. The input energy spectra, $P(E)$, may come from different configurations of two interacting biomolecules, or they can come
from a single biomolecule interacting with a set of partner binders. It is important to note that our conclusions hold true even when the probability distribution, $P(E_1)$
with a larger standard deviation than $P(E_2)$, $\sigma_1>\sigma_2$, is sampled by a smaller number of states, $M_1<M_2$!
The free energy $F_1$ will be reduced compared with $F_2$ in the latter case due to the fact that the dominant contribution
to the partition function comes from the lower-energy tails of $P(E_1)$ and $P(E_2)$, and thus a wider energy spectrum will always deliver 
a lower free energy. 

In conclusion, we stress that our model is applicable to weakly interacting systems, without a pronounced energy minima in the interaction energy spectrum.
We use the term ``nonspecific binding'' or ``promiscuous binding'' to describe such systems. Nonspecific binding is widespread in a living cell.
In practice, the majority of the interactions are actually nonspecific. Traditionally, such nonspecific interactions are neglected, which leads
to significant inaccuracies in the computation of the free energy of the system. Here, we suggested a possible method to estimate the relative
free energies of nonspecific binding. 
\\
\\
We thank A. Afek, D. Andelman, D. Frenkel, O. Furman (Schueler), W. Gelbart, and E. I. Shakhnovich for helpful discussions. D. B. L. acknowledges the financial support from the Israel Science Foundation grant 1014/09.



\begin{thebibliography}{1}



\bibitem{ppi}
Yu, H., Braun, P.,  Yildirim, M.A., Lemmens, I., Venkatesan, K.,
{\it{et al}.}.:
High-quality binary protein interaction map of the yeast interactome network.
{\it{Science}} {\bf{322}}, 104-110 (2008). doi: 10.1126/science.1158684

\bibitem{polly} 
Fordyce, P.~M., Gerber, D., Tran, D., Zheng, J., Li, H., {\it{et al.}}.:
{\it{De novo}} identification and biophysical characterization of transcription-factor binding sites with microfluidic affinity analysis.
{\it{Nature Biotech.}} {\bf{28}}, 970-975 (2010).

\bibitem{rna}
Hafner, M., Landthaler, M., Burger, L., Khorshid, M., Hausser, L.,
{\it{etal.}}.:
Transcriptome-wide Identification of RNA-Binding Protein and MicroRNA Target Sites by PAR-CLIP.
{\it{Cell}}  {\bf{141}}, 129-141 (2010). doi: 10.1016/j.cell.2010.03.009
    
\bibitem{lukatsky_prl}
Lukatsky, D.~B., Zeldovich, K.~B., Shakhnovich, E.~I.: 
Statistically Enhanced Self-Attraction of Random Patterns.
{\it{Phys. Rev. Lett.}} {\bf{97}}(17), 178101 (2006). doi: 10.1103/PhysRevLett.97.178101
    
\bibitem{lukatsky_jmb} 
Lukatsky, D.~B., Shakhnovich, B.~E., Mintseris, J., Shakhnovich, E.~I.: 
Structural Similarity Enhances Interaction Propensity of Proteins.
{\it{J. Mol. Biol.}} {\bf{365}}, 1596-1606 (2007). doi:10.1016/j.jmb.2006.11.020
    

\bibitem{baker_pnas}
Andre, I., Strauss, C.~E.~M., Kaplan, D.~B., Bradley, P., Baker, D.:
Emergence of symmetry in homooligomeric biological assemblies.
{\it{Proc. Natl. Acad. Sci. U.S.A.}} {\bf{105}}, 16148-16152 (2008). doi: 10.1073/pnas.0807576105

\bibitem{lukatsky_pre} 
Lukatsky, D.~B., Shakhnovich, E.~I.:
Statistically enhanced promiscuity of structurally correlated patterns
{\it{Phys. Rev. E.}} {\bf{77}}, 020901(R) (2008). doi: 10.1103/PhysRevE.77.020901

\bibitem{lukatsky_ariel}
Lukatsky, D.~B., Afek, A., Shakhnovich, E.~I.:
Sequence correlations shape protein promiscuity. J. Chem. Phys. {\bf{135}}(6), 065104 (2011). doi:10.1063/1.3624332

\bibitem{lukatsky_jmb_2011}
Afek, A., Shakhnovich, E.~I.,  Lukatsky, D.~B.:
Multi-scale sequence correlations increase proteome structural disorder and promiscuity.
J. Mol. Biol. {\bf{409}}(3), 439-449 (2011). doi: 10.1016/j.jmb.2011.03.056

\bibitem{sela_2011}
Sela, I.,  Lukatsky, D.~B.:
DNA sequence correlations shape nonspecific transcription factor-DNA binding affinity.
Biophys. J. {\bf{101}}(1),160-166 (2011). doi: 10.1016/j.bpj.2011.04.037


\bibitem{derrida}
Derrida, B.: 
Random-Energy Model: Limit of a Family of Disordered Models.
{\it{Phys. Rev. Lett.}} {\bf{45}}(2), 79-82 (1980). doi: 10.1103/PhysRevLett.45.79


\bibitem{tawfik}
Khersonsky, O., Tawfik, D. S.: 
Enzyme promiscuity: a mechanistic and evolutionary perspective.
Annu. Rev. Biochem. {\bf{79}}, 471-505 (2010). doi: 10.1146/annurev-biochem-030409-143718

\bibitem{AS08}
Alon, N. \& Spencer, J.,
\newblock {\em The probabilistic method}.
\newblock Wiley-Interscience Series in Discrete Mathematics and Optimization,
  3rd edition, 2008.

\bibitem{AB57}
Atchinson, J. and Brown, J.~A.~C.,
\newblock {\em The Lognormal Distribution, with Special Reference to Its Use in
  Economics}.
\newblock Cambridge University Press, 1957.

\bibitem{shakh}
Shakhnovich, E.~I., Gutin, A.~M.:
Formation of unique structure in polypeptide chains: Theoretical investigation with the aid of a replica approach.
{\it{Biophys. Chem.}} {\bf{34}}(3), 187-199 (1989).


\end{thebibliography}

\end{document}